\documentclass[runningheads]{llncs}
\usepackage[T1]{fontenc}
\usepackage{graphicx}
\usepackage{amsmath} 
\usepackage{xspace}
\usepackage{booktabs}
\usepackage{hyperref}
\hypersetup{hidelinks}
\usepackage{array}
\usepackage{todonotes}
\RequirePackage[capitalise,nosort]{cleveref}
\usepackage{tikz}
\usetikzlibrary{arrows, positioning, backgrounds, fit, calc, shapes.multipart}

\usepackage{listings}
\usepackage{algorithm}
\usepackage[noend]{algpseudocode}

\usepackage{caption}

\usepackage{wrapfig}
\usepackage{newfloat}
\DeclareFloatingEnvironment[
  fileext=c,
  name=Listing,
]{listing}

\usepackage{subcaption}
\DeclareCaptionSubType{listing}

\newcommand{\fizzer}{\textsc{Fizzer}\xspace}

\newcommand{\abe}{\textsc{abe}\xspace}

\renewcommand{\orcidID}[1]{{\href{https://orcid.org/#1}{\protect\raisebox{3.25pt}{\protect\includegraphics{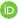}}}}}

\newcolumntype{L}[1]{>{\raggedright\let\newline\\
\arraybackslash\hspace{0pt}}m{#1}}
\newcolumntype{C}[1]{>{\centering\let\newline\\
\arraybackslash\hspace{0pt}}m{#1}}
\newcolumntype{R}[1]{>{\raggedleft\let\newline\\
\arraybackslash\hspace{0pt}}m{#1}}
\newcolumntype{P}[1]{>{\raggedright\tabularxbackslash}p{#1}}

\begin{document}
%
\title{Gray-Box Fuzzing in Local Space\thanks{This work has been supported by the Czech Science Foundation grant GA23-06506S.}}
%
%
\author{Martin Jon\'{a}\v{s} \orcidID{0000-0003-4703-0795}
  \and
  Jan Strej\v cek \orcidID{0000-0001-5873-403X}
  \and
  Marek Trt\'ik \orcidID{0009-0009-6122-9574}
}
\authorrunning{M. Jonáš et al.}
%
\institute{
  Faculty of Informatics,
  Masaryk University,
  Brno,
  Czech Republic\\
  \email{\{martin.jonas,strejcek,trtikm\}@mail.muni.cz}
}
%
\maketitle              
\begin{abstract}
  We consider gray-box fuzzing of a program instrumented such that information
  about evaluation of program expressions converting values of numerical types
  to Boolean, like $x <= y$, is recorded during each program's execution. Given
  that information for an executed program path, we formally define the problem
  for finding input such that program's execution with that input evaluates all
  those expressions in the same order and with the same Boolean values as in the
  original execution path, except for the last one, which is evaluated to the
  opposite value. Then we also provide an algorithm searching for a solution of
  the problem effectively. The effectiveness of the algorithm is demonstrated
  empirically via its evaluation on the TestComp 2024 benchmark suite.
\end{abstract}

\section{Introduction}
\label{sec:intro}

The primary goal of gray-box fuzzing~\cite{fuzzing18} is to find as many defects
in an analyzed program as possible. More precisely, the goal is to find
\emph{inputs} to the program such that program's execution with them triggers
different defects in the program.

In this paper we use a simplified model for input. The program reads input via
calls to reserved functions returning a value of a requested primitive type,
like \texttt{char}, \texttt{int}, \texttt{float}, etc. For example, by calling
\texttt{read\_int()} the program receives an input value of the type
\texttt{int}. Each call to these functions during program's execution creates a
fresh \emph{input variable} of the corresponding type holding the returned
value. The input variables are artificial, i.e., they do not appear in the
program. They just represent the input for which the program executes along the
path where the input variables were created.

During gray-box fuzzing we do not read or otherwise use program's code; neither
source code nor binary. However, we record information about each program's
execution via probing code, which we instrument to the program before fuzzing is
started. The recorded information is used for generation of next inputs.

Execution of the probes slows down overall program's execution. If we instrument
too much or too complex probes to the program, then the number of executions
within a given time budget can drop considerably compared to number of
executions of the original (not instrumented) program. In that case we in fact
decrease our chances to discover a defect. If the probes do not give us enough
information for next input generation, we can easily generate inputs for which
the program executes along paths executed before. We thus also decrease our
chances to discover a defect. An optimal approach is to instrument as few and as
simple probes as possible, which still provide enough information for generation
of inputs for which program executes new paths.


In this paper we consider a gray-box fuzzing, where probes record information
about evaluation of program expressions converting values of numerical types to
Boolean, like $x <= y$, $x == 5$. That is because Boolean values are primarily
involved in program branching. We call any such expression as \emph{atomic
Boolean expression} (\abe). We give precise definition later. For each executed
\abe we record the Boolean value it was evaluated to and also a signed distance
to evaluation of the \abe to the opposite value. For example, signed distance
function for a comparison is the difference between the left and right operand,
e.g., for $x <= y - 5$ the signed distance is $x - y + 5$. But both operands
must be casted to a wider type to prevent an overflow/underflow when taking
their difference.

So, for each executed program path we have a list of input variables, the
sequence of \abe{}s in the order of their execution and for each \abe{} we know
its Boolean value and the signed distance value. In the worst case, we can
assume that all input variables affect signed distance of each \abe{}. In our
implementation, we apply taint-flow analysis along that execution path to
optimize the correspondence between input variables and signed distances of
\abe{}s.

The contribution of the paper is the following. Given all mentioned information
associated with an executed program path, we first formally define the problem
for finding input such that program's execution with that input evaluates all
\abe{}s in the same order and with the same Boolean values as in the original
execution path, except for the last \abe{}, which is evaluated to the opposite
value. That is done in Section~\ref{sec:definition}. Then we describe an
algorithm searching for a solution of the problem effectively. That is done in
Sections~\ref{sec:embedding_valuations}-\ref{sec:implementation}. Finally,
we demonstrate the effectiveness of the algorithm by its evaluation on the
TestComp 2024 benchmark suite; we show improvement in outcomes over those from
the original version of the gray-box fuzzer, called \fizzer, into which we
implemented this algorithm. That is done in Section~\ref{sec:evaluation}.

\begin{remark}
  \label{rem:copverage}
  When the algorithm finds a solution to the problem, then we have two inputs
  (from the given path and the found one). They lead to evaluation of the last
  \abe{} to opposite Boolean values. We thus say that \abe{} is covered. That is
  the reason for calling the problem as \emph{coverage problem}. Further note
  that if we are given information for an executed program path, then we can
  formulate the coverage problem for any prefix of that path.
\end{remark}

\section{Coverage problem definition}
\label{sec:definition}

Since we intensively use vectors and vector spaces in the rest of the paper, we
start with the basic notation. If $\vec{u}$ is a vector, then $\vec{u}_i$ is the
$i$-th coordinate of $\vec{u}$. If $\vec{u}$ has all coordinates 0 except
$\vec{u}_i$ which is 1, then we call $\vec{u}$ as $i$-th axis. If $\vec{u}$ is a
vector of vectors, then $\vec{u}_{i,j}$ denotes the $j$-th coordinate of the
$i$-th vector of $\vec{u}$. We naturally extend this notation to arbitrary
nesting depth of vectors. We use only column vectors in this paper. Lastly, we
denote the dot product of vectors $\vec{u},\vec{v}$ as $\vec{u} \cdot \vec{v}$
and the length of a vector $\vec{u}$ as $|\vec{u}|$. Note that $|\vec{u}| =
\sqrt{\vec{u} \cdot \vec{u}}$.

A \emph{type} is any signed or unsigned 8-, 16-, 32- or 64-bit integer or 32- or
64-bit floating point type of C language. The Boolean type is modelled as 8-bit
unsigned integer with only two possible values 0 and 1. We denote a type of a
variable $x$ as $\tau(x)$ and the finite size of the set of all possible values
of the type as $|\tau(x)|$. We use the symbols $x$ or $x'$ possibly with the
lower index for denoting variables.

An \emph{atomic Boolean expression} (\abe) is a function computing a value of
the Boolean type from at least one parameter of a numerical type such that no
logical connective is involved in the computation and also there is no \abe
comprised within definition of the function. The comparison instructions, like
$x_1 = x_2$, $x_1 \geq 10$ etc., represent the most common and also the most
important \abe{}s. The type casting or truncating of a numerical value to a
Boolean value are another examples of \abe{}s.

A \emph{valuation} is any function $I$ assigning each variable $x \in Dom(I)$ a
value $I(x) \in |\tau(x)|$.

A \emph{comparator} is any element of the set $\{=,\neq,<,\leq,>,\geq\}$. If $P$
is a comparator and $a,b$ values of the same type, then $P(a,b)$ forms a
predicate. The \emph{opposite} comparator to $=,\neq,<,\leq,>,\geq$ is
$\neq,=,\geq,>,\leq,<$, respectively.

A \emph{black-box function} $F$ is a function from input variables, denoted as
$\nu(F)$ and called \emph{parameters}, to a 64-bit floating point number. The
algorithm of $F$ is not available to us, but given a valuation $I$, we can call
it with values $I(\nu(F))$ for the parameters to obtain the resulting value
$F[I]$. However, the call \emph{fails} to compute $F[I]$, if $I(\nu(F)) \not\in
Dom(F)$. Whenever we use the value $F[I]$, e.g., in some expression, we
automatically assume the call to $F$ with values $I(\nu(F))$ did not fail. We
use black-box functions for signed distances of \abe{}s.

Let $\vec{F}$ be a vector of $n$ black-box functions. A valuation $I$ is a
valuation of $\vec{F}$, if it is defined for all variables $\nu(\vec{F}) =
\nu(\vec{F}_1) \cup \cdots \cup \nu(\vec{F}_n)$. Without loss of generality we
assume $\nu(\vec{F}) = \{x_1,\ldots,x_{|\nu(\vec{F})|}\}$.

Let $\vec{F}$ be a vector of $n$ black-box functions such that $\nu(\vec{F}_n)
\neq \emptyset$, $\vec{P}$ be a vector of a dimension $n$ of comparators and $I$
be a valuation of $\vec{F}$ such that calls to functions
$\vec{F}_1,\ldots,\vec{F}_n$ in that order yield values
$\vec{F}_1[I],\ldots,\vec{F}_n[I]$ for which we get $\vec{P}_i(\vec{F}_i[I], 0)$
is true for all $1 \leq i < n$ and $\vec{P}_n(\vec{F}_n[I], 0)$ is false. Then
the tuple $\mathcal{S}=(\vec{F},\vec{P}, I)$ is a \emph{coverage problem} of
size $n$.

Let $\mathcal{S}=(\vec{F},\vec{P}, I)$ be a coverage problem of size $n$ and $J$
be a valuation. The order in which we can call the black-box functions in the
vector $\vec{F}$ is fixed. Namely, $\vec{F}_i$ is called before $\vec{F}_j$, if
$i < j$. Furthermore, if the call of $\vec{F}_i$ fails or
$\vec{P}_i(\vec{F}_i[J], 0)$ is false, then we cannot call any function
$\vec{F}_j$, where $j > i$. Based on that we can introduce the following
convention. Whenever we use the value $\vec{F}_i[J]$, e.g., in
$\vec{P}_i(\vec{F}_i[J], 0)$, we automatically assume that
$\vec{P}_j(\vec{F}_j[J], 0)$ is true for all $j < i$. If
$\vec{P}_n(\vec{F}_n[J], 0)$ is true, then $J$ is a \emph{solution} of
$\mathcal{S}$.

We typically build a coverage problem from an executed program path. We
represent the path by a vector of \abe{}s evaluated along that path. We call the
vector as a \emph{trace}. Each \abe{} evaluated to false is present in the trace
with the opposite comparator, except the last \abe{}, which is present in the
trace with the opposite comparator, if it is evaluated to true. Then,
$\vec{F}_i$ and $\vec{P}_i$ are the signed distance and the comparator of the
$i$-th \abe{} in the trace, respectively. The valuation $I$ represents the input
variables created along the path.

\begin{example}
  \label{ex:1}
  Let us consider a trace $(x_1 = x_2, x_1 \geq 10)$, where $x_1,x_2$ are input
  variables of 64-bit floating point type. Then we have
  $\vec{F}=([x_1-x_2],[x_1-10])$\footnote{The square brackets declare the
  computation (algorithm) is in fact hidden.}, and $\vec{P}=(=,\geq)$. If we
  also include a valuation $I$ mapping both variables to zero, then we can
  easily check that $\mathcal{S}=(\vec{F},\vec{P}, I)$ is a coverage problem of
  size 2. Clearly, a valuation mapping both $x_1,x_2$ to 10 is one possible
  solution to $\mathcal{S}$. On the other hand, for a valuation $J$ mapping
  $x_1,x_2$ to values 0,1 we get $\vec{F}_1[J]=-1$ and $\vec{P}_1(\vec{F}_1[J],
  0)$ equal to false, which makes it impossible for us to call $\vec{F}_2$.
\end{example}


For a coverage problem $\mathcal{S}=(\vec{F},\vec{P}, I)$ of size $n$ we can
construct a set $D \subseteq \nu(\vec{F})$ as follows. We initialize $D$ to
$\nu(\vec{F}_n)$. Then, while there exist $1\leq i < n$ such that $D$ contains
at least one variable in $\nu(\vec{F}_i)$, but not all of them, then we insert
all the variables in $\nu(\vec{F}_i)$ to $D$. The coverage problem $\mathcal{S}$
is called \emph{reduced}, if for all $1\leq i < n$ we have $D \cap
\nu(\vec{F}_i) \neq \emptyset$. We can easily check that the coverage problem in
the example above is reduced ($D=\{x_1,x_2\}$).


If the coverage problem $\mathcal{S}=(\vec{F},\vec{P}, I)$ of size $n$ is not
reduced, then we can use the set $D$ to construct a reduced coverage problem
$\mathcal{S}'=(\vec{F}',\vec{P}', I')$ as follows. We initialize all components
of $\mathcal{S}'$ to be equal to the corresponding components of $\mathcal{S}$.
Next, we remove each $i$-th coordinate from $\vec{F}'$ and $\vec{P}'$, where $D
\cap \nu(\vec{F}_i) = \emptyset$. Lastly, we restrict $I'$ to $D$. So, if $J'$
is a solution of the reduced coverage problem $\mathcal{S}'$, then a valuation
$J$ returning for each $x \in \nu(\vec{F}')$ value $J'(x)$ and for any other
variable $x'$ value $I(x')$ is a solution of the original non-reduced coverage
problem $\mathcal{S}$. Because of that, we assume coverage problems are reduced
in the rest of the paper, unless explicitly said otherwise.

\begin{example}
  Let us consider a trace $(x_1=x_2, x_3 \geq 10)$, where all variables are of
  64-bit floating point type. Then we have $\vec{F}=([x_1-x_2],[x_3-10])$ and
  $\vec{P}=(=,\geq)$. With a valuation $I$ mapping $x_1,x_2,x_3$ to zero we get
  a coverage problem $\mathcal{S}=(\vec{F},\vec{P}, I)$ of size 2 which is not
  reduced, because $D=\{x_3\}$ and $\nu(\vec{F}_1)=\{x_1,x_2\}$. Using $D$ we
  compute $\vec{F}'=([x_3-10])$, $\vec{P}'=(\geq)$ and with a valuation $I'$
  mapping $x_3$ to zero we get a reduced coverage problem
  $\mathcal{S}'=(\vec{F}',\vec{P}', I')$. A possible solution of $\mathcal{S}'$
  is a valuation $J'$ mapping $x_3$ to 10. A valuation $J$ extending $J'$ for
  $x_1,x_2$ according to $I$ (i.e., $x_1=x_2=0$) is then a solution of
  $\mathcal{S}$.
\end{example}

\begin{remark}
  In examples above we defined the (hidden) algorithm of each black-box function
  directly from the corresponding \abe{}. In practice, the algorithm may consist
  of other instructions of the program executed along the path. For instance,
  consider a path consisting of instructions $x_2 := 2*x_1-1$ and $x_2=5$. There
  is only one \abe, $x_2=5$, in this path. The algorithm of the corresponding
  black-box function is $[2*x_1-6]$ (not $x_2 - 5$) and $x_1$ is function's
  parameter (not $x_2$). However, for simplicity of presentation, black-box
  functions in all our examples can directly be inferred from the corresponding
  \abe{}s.
\end{remark}

\section{Embedding valuations to a vector space}
\label{sec:embedding_valuations}

Let $\mathcal{S}=(\vec{F},\vec{P}, I)$ be a coverage problem of the size $n$.

We can embed all possible valuations of the variables $\nu(\vec{F})=
\{x_1,\ldots,x_{|\nu(\vec{F})|}\}$ into the arithmetic vector space over the
field of real numbers\footnote{In our implementation we model real numbers by
64-bit floating points.} of the dimension $|\nu(\vec{F})|$ with the dot product
as follows.

The space certainly has the orthonormal basis $B$, where for each $1\leq i \leq
|\nu(\vec{F})|$ the $i$-th basis vector $B_i$ is the $i$-th axis. We denote a
vector space generated by (any basis) $B$ as $[B]$. 

Let $J$ be a valuation of the variables $\nu(\vec{F})$. All values assigned to
variables by $J$ can directly be expressed by the corresponding real numbers. We
can thus express $J$ as a linear combination of basis vectors, i.e., as the
vector $\vec{J} = \sum_{i=1}^{|\nu(\vec{F})|}x_i B_i =
(x_1,\ldots,x_{|\nu(\vec{F})|})^\top \in [B]$.

Conversely, given a vector $\vec{J} \in [B]$ we can compute the valuation $J$
from $\vec{J}$ such that $J(x_i)$ is a value $v \in \tau(x_i)$ such that $|v -
\vec{J}_i|$ is the smallest.

\begin{example}
  Let us consider a trace $(x_1=x_2)$, where $x_1,x_2$ are of 32-bit signed
  integer type and $I$ be a valuation mapping mapping $x_1,x_2$ to values 2,-3.
  Then $\mathcal{S}=(\vec{F},\vec{P}, I)$, where $\vec{F}=([x_1-x_2])$ and
  $\vec{P}=(=)$, is a coverage problem. The basis $B = \{ (1,0)^\top, (0,1)^\top
  \}$. We can encode $I$ using a vector
  $\vec{I}=2B_1+(-3)B_2=2(1,0)^\top-3(0,1)^\top=(2,-3)^\top$. Conversely, a
  vector $\vec{J}=-1.23B_1+2.7B_2=(-1.23,2.7)^\top$ gives us a valuation $J$
  assigning -1 to $x_1$ and 3 to $x_2$.
\end{example}

Observe that if $\vec{I} \in [B]$ represents a valuation $I$, then a vector
$\vec{I} + \vec{u}$, where $\vec{u} \in [B]$ is any vector, also represents some
valuation.

In the rest of the paper valuations will be represented via the corresponding
vectors. Their conversion to actual valuations is implicit. This convention
allows us, for example, to say: We call functions in $\vec{F}$ with the vector
$\vec{I} + \vec{u}$. We can also write $\vec{F}_i[\vec{I} + \vec{u}]$, and so
on. Lastly, if $I$ denotes a valuation, then $\vec{I}$ automatically denotes the
corresponding vector.

\section{Local spaces}
\label{sec:local_spaces}

Let $\mathcal{S}=(\vec{F},\vec{P}, I)$ be a coverage problem of the size
$n$.

The goal of this section is to compute a vector $\vec{B}$ of orthonormal bases,
one for each black-box function in $\vec{F}$. $\vec{B}$ thus comprises $n$
orthonormal bases, where $\vec{B}_i$ corresponds to $\vec{F}_i$. Each $\vec{B}_i
= \{ \vec{B}_{i,1}, \ldots, \vec{B}_{i,|\vec{B}_i|}\}$ is an orthonormal basis
of the arithmetic vector space over the field of real numbers with the dot
product. More Specifically, $\vec{B}_1$ is exactly the basis $B$ discussed in
the section~\ref{sec:embedding_valuations}. For all $2 \leq i \leq n$, each
basis vector $\vec{B}_{i,j} \in \vec{B}_i$ is a vector from the vector space
$[\vec{B}_{i-1}]$. That is, $\vec{B}_{i,j} =
(\vec{B}_{i,j,1},\ldots,\vec{B}_{i,j,|\vec{B}_{i-1}|})^\top$ is a linear
combination of all basis sectors in $\vec{B}_{i-1}$, i.e., $\vec{B}_{i,j} =
\sum_{k=1}^{|\vec{B}_{i-1}|}\vec{B}_{i,j,k}\vec{B}_{i-1,k}$.

Observe, that due to the recurrent nature of the definition, $\vec{B}_{i,j}$ can
also be expressed as a linear combination of basis sectors in $\vec{B}_{i-2}$
(if $i-1 \geq 2$), because each vector $\vec{B}_{i-1,k}$ is a linear combination
of vectors in $\vec{B}_{i-2}$. Therefore, $\vec{B}_{i,j}$ can eventually be
expressed as a linear combination of basis vectors in $\vec{B}_1$. That further
implies that if $\vec{u}$ is a linear combination of basis vectors in
$\vec{B}_i$, then it can also be expressed as a linear combination of basis
vectors in $\vec{B}_1$, which we denote as $\vec{u}^*$.

We call the vector $\vec{B}$ as the vector of \emph{local spaces} of black-box
functions $\vec{F}$, where $\vec{B}_i$ is called the \emph{local space} of
$\vec{F}_i$.

The purpose of the vector $\vec{B}$ is to boost the search for a solution of
$\mathcal{S}$. Of course, the bases in $\vec{B}$ must be computed in a certain
way. We start the description of their computation with a motivation example.

\begin{figure*}[t]
  \begin{tabular}{ccc}
    \includegraphics[width=3.5cm]{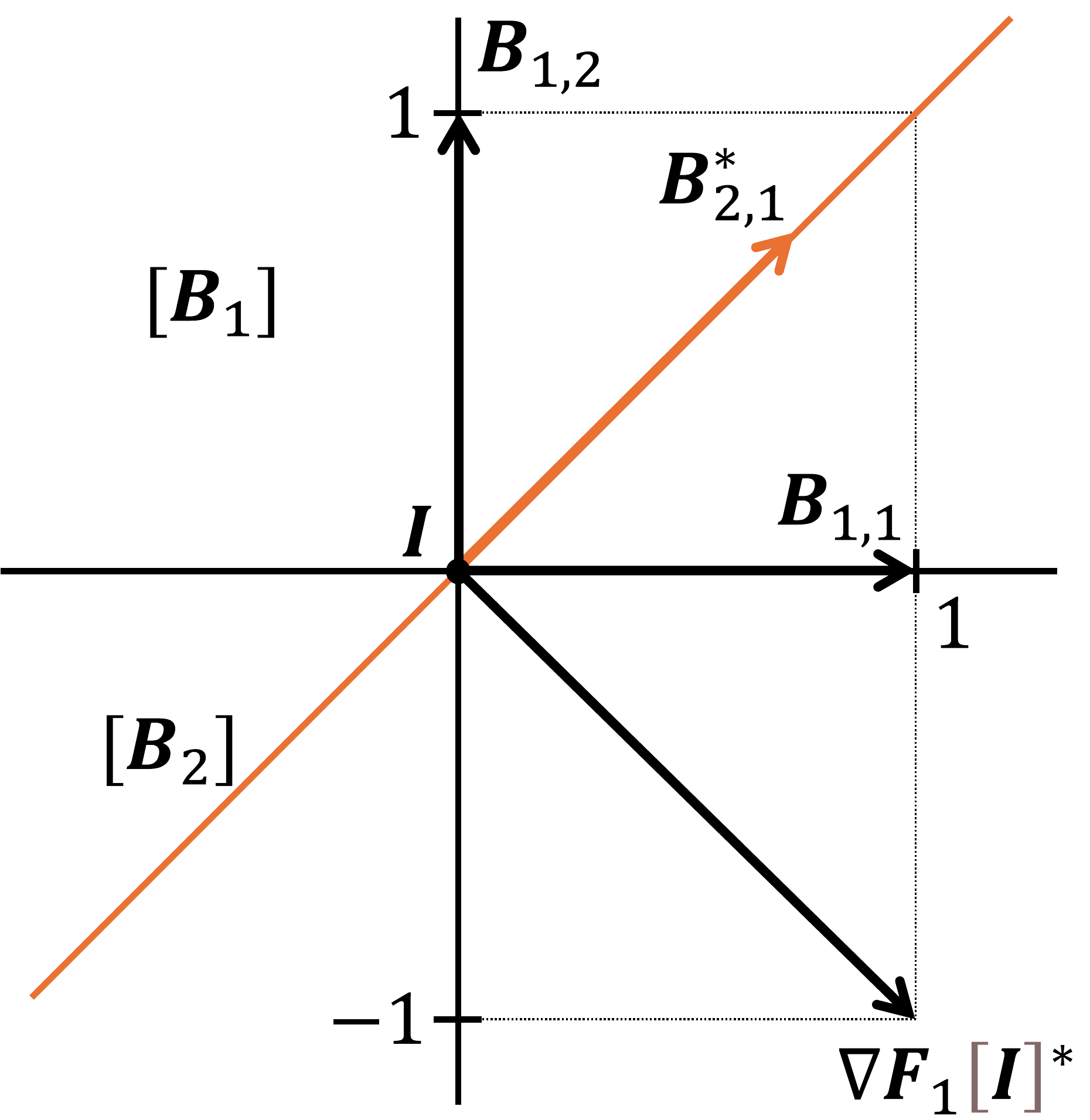} &
      \includegraphics[width=3.5cm]{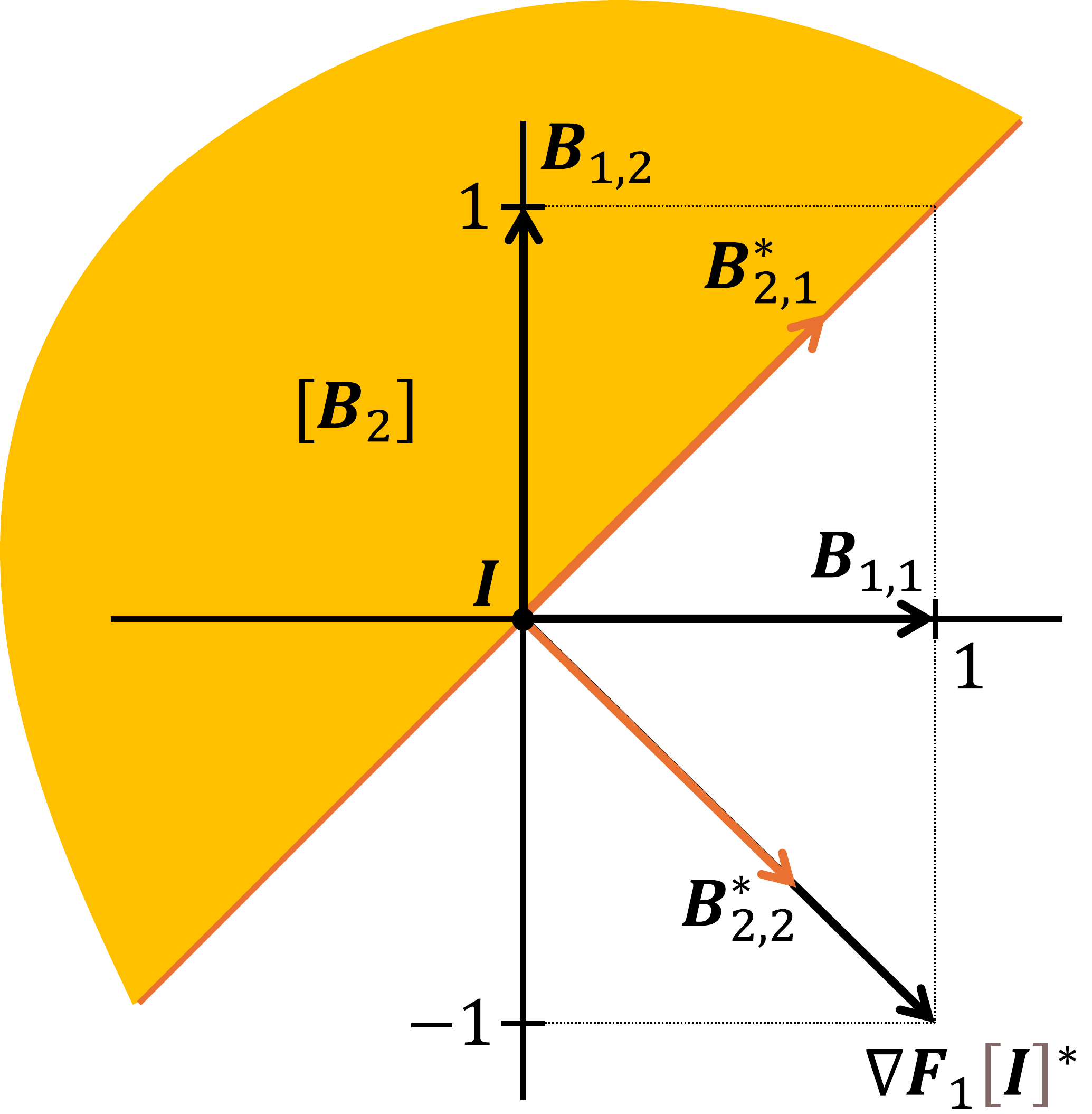} &
      \includegraphics[width=4cm]{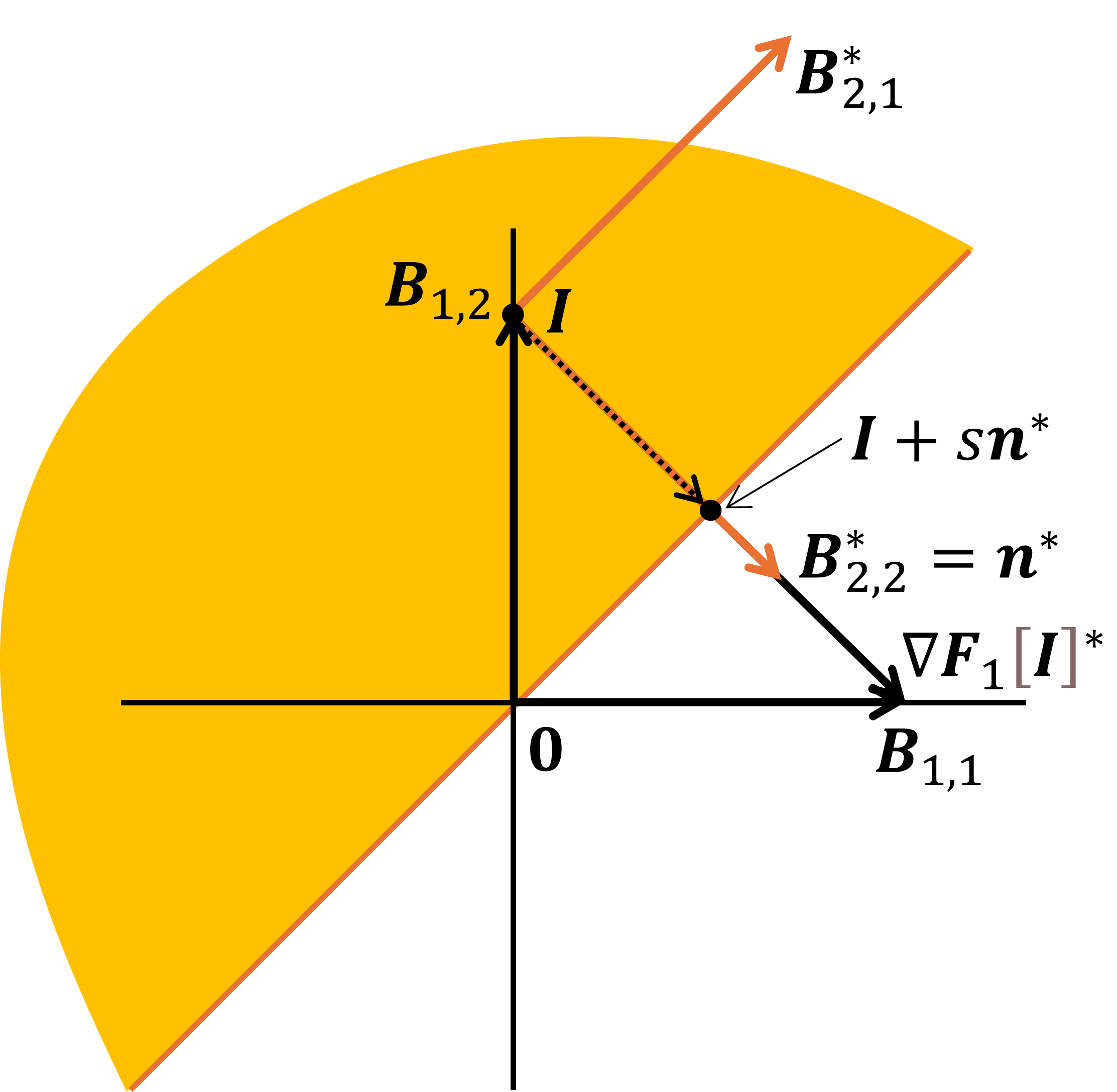} \\
      (a) & (b) & (c) \\
    \includegraphics[width=4cm]{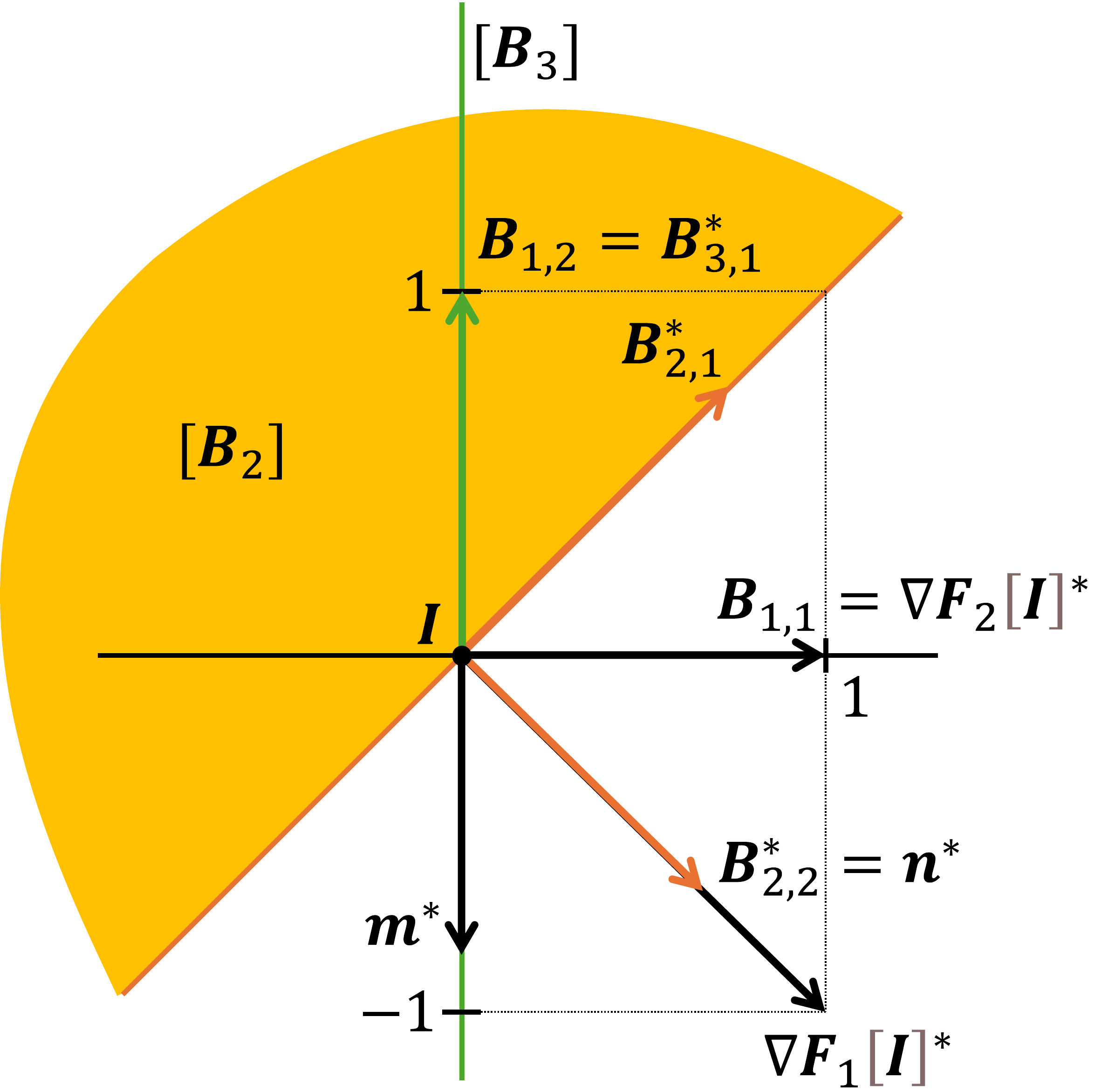} &
      \includegraphics[width=3.5cm]{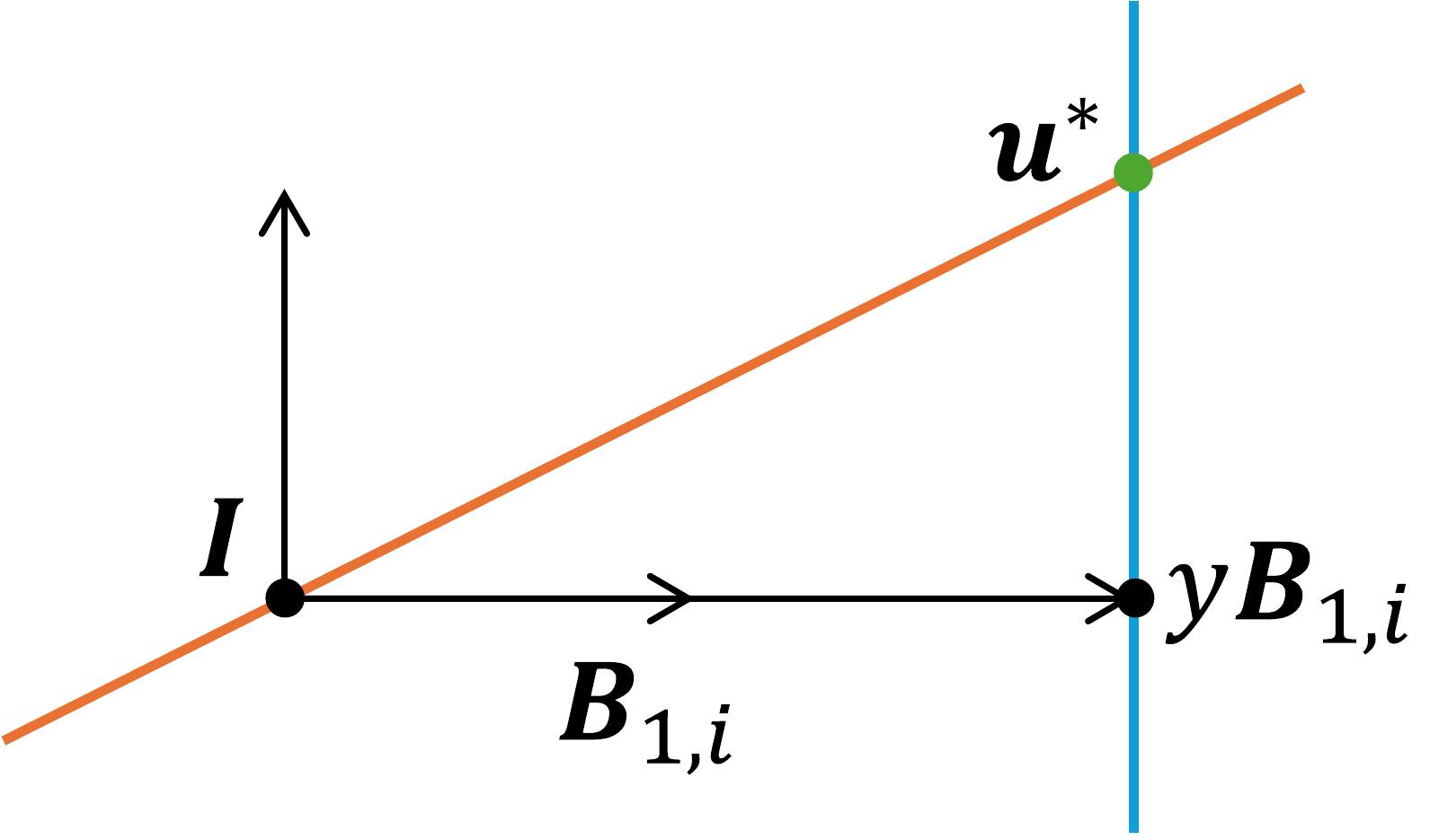} &
      \includegraphics[width=4cm]{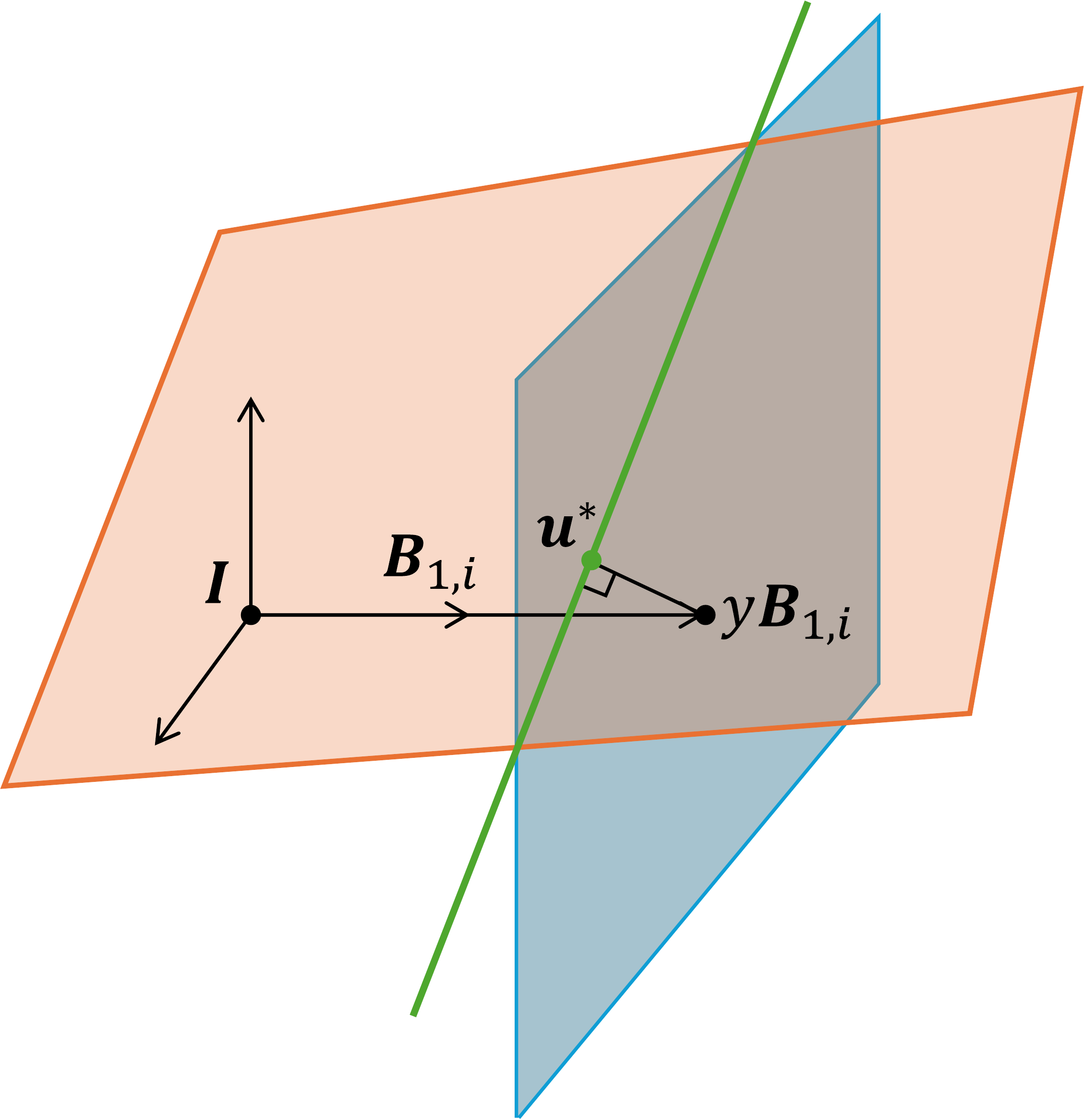} \\
      (d) & (e) & (f)
  \end{tabular}
  \caption{Pictures used in examples.}
  \label{fig:examples}
\end{figure*}

\begin{example}
  \label{ex:local_space:1}
  Let us consider the coverage problem $\mathcal{S}=(\vec{F},\vec{P}, I)$ of
  size 2 defined in Example~\ref{ex:1}. The vector $\vec{B}=(\vec{B}_1,
  \vec{B}_2)^\top$, where $\vec{B}_1 = \{ \vec{B}_{1,1}=(1,0)^\top,
  \vec{B}_{1,2}=(0,1)^\top \}$. If we construct $\vec{B}_2$ such that for each
  $\vec{u} \in [\vec{B}_2]$ the predicate
  $\vec{P}_1(\vec{F}_1[\vec{I}+\vec{u}^*],0)$ evaluates to true, then the search
  for a solution of $\mathcal{S}$ will be simpler than without $\vec{B}$,
  because we can now focus on satisfying only
  $\vec{P}_2(\vec{F}_2[\vec{I}+\vec{u}^*],0)$.

  The construction of $\vec{B}_2$ is based on finding a set of all vectors
  $\vec{u} \in [\vec{B}_1]$ such that $\vec{F}_1[\vec{I} + \vec{u}^*] =
  \vec{F}_1[\vec{I}]$. In general, the set does not have to form a vector space.
  However, we can approximate it by a vector space orthogonal to the gradient of
  $\vec{F}_1$ at the vector $\vec{I}$.

  In this example we compute the gradient from the hidden code of $\vec{F}_1$.
  Later we describe the proper numerical computation. So, we have $\nabla
  \vec{F}_1[\vec{I}] = (1,-1)^\top \in [\vec{B}_1]$. The subspace of
  $[\vec{B}_1]$ orthogonal to the gradient at $\vec{I}$ forms the line $\vec{I}
  + t(1,1)^\top$, where $t$ is any real number. We thus define $\vec{B}_2 = \{
  \vec{B}_{2,1} = \frac{1}{\sqrt{2}}(1,1)^\top\}$. This is graphically depicted
  in the Fig.~\ref{fig:examples}(a).

  With the basis $\vec{B}_2$ computed, we can search for a solution of
  $\mathcal{S}$ in one dimensional space $[\vec{B}_2]$. Namely, we search for a
  vector $\vec{v} = (t)^\top \in [\vec{B}_2]$ such that
  $\vec{P}_2(\vec{F}_2[\vec{I}+\vec{v}^*],0)$ is true, where $\vec{v}^* =
  {(t)^\top}^* = t \vec{B}_{2,1} = t(\frac{1}{\sqrt{2}}\vec{B}_{1,1} +
  \frac{1}{\sqrt{2}}\vec{B}_{1,2}) = t(\frac{1}{\sqrt{2}}(1,0)^\top +
  \frac{1}{\sqrt{2}}(0,1)^\top) = \frac{t}{\sqrt{2}}(1,1)^\top$. Clearly, for
  $t=10\sqrt{2}$ we get a solution of $\mathcal{S}$. We postpone the definition
  of the algorithm computing a solution to Section~\ref{sec:searching}.
\end{example}

Let us suppose we already have an orthonormal basis $\vec{B}_i$ computed, where
$1 \leq i < n$. Our goal is to compute the orthonormal basis $\vec{B}_{i+1}$.

For that we must first compute the gradient $\nabla\vec{F}_i[\vec{I}] \in
[\vec{B}_i]$. For each $1 \leq j \leq |\vec{B}_i|$ we approximate the $j$-th
partial derivative $\nabla_j\vec{F}_i[\vec{I}]$ using a finite difference
$$
  \nabla_j\vec{F}_i[\vec{I}] = \frac{\vec{F}_i[\vec{I}+\epsilon\vec{e}_j^*] -
  \vec{F}_i[\vec{I}]}{\epsilon}
$$
where $\vec{e}_j \in [\vec{B}_i]$ is the $j$-th axis vector, and the number
$\epsilon$ represents a step along the line $\vec{I}+\epsilon\vec{e}_j^*$, which
is computed according to algorithm in Section~\ref{sec:line_epsilon_step}.
This algorithm is started with an initial value for $\epsilon$, computed by
algorithm in Section~\ref{sec:epsilon_step}, from the value $\max \{
|\vec{I}_j|~;~1\leq j \leq \nu(\vec{F}) \}$. If the call of $\vec{F}_i$ for the
valuation $\vec{I}+\epsilon\vec{e}_j^*$ fails, then we try to compute
$\nabla_j\vec{F}_i[\vec{I}]$ for $-\epsilon$. If this attempt fails too, then we
set $\nabla_j\vec{F}_i[\vec{I}]$ to zero. Of course, this may lead to a
reduction of performance boost we expect from the resulting vector $\vec{B}$.


With the gradient $\nabla\vec{F}_i[\vec{I}]$ available we can compute individual
vectors of $\vec{B}_{i+1}$. If $|\nabla\vec{F}_i[\vec{I}]|=0$, then $\vec{F}_i$
is constant. That means we want $[\vec{B}_{i+1}] = [\vec{B}_i]$. So, for each $1
\leq j \leq |\vec{B}_i|$ we set the vector $\vec{B}_{i+1,j}$ as the $j$-th axis.
Otherwise, we compute $\vec{B}_{i+1,j}$ as depicted in
Algorithm~\ref{alg:compute_next_basis}. The computation of a vector $\vec{w}$ to
be (normalized and) added to $\vec{B}_{i+1}$ is based on the following idea. We
initialize $\vec{w}$ as an axis vector. Then we remove from it all components
which are parallel to any vector listed at
line~\ref{alg:compute_next_basis:enum_vectors}. The vector $\vec{w}$ at
line~\ref{alg:compute_next_basis:w} is then orthogonal to all mentioned vectors.

\begin{algorithm}
  \caption{Computes $\vec{B}_{i+1}$ from $\nabla\vec{F}_i[\vec{I}]$ and $\vec{B}_i$,
    where $|\nabla\vec{F}_i[\vec{I}]| \neq 0$}
    \label{alg:compute_next_basis}
  \begin{algorithmic}[1]
    \State $\vec{B}_{i+1} := \emptyset$
    \ForAll{$j=1,\ldots,|\vec{B}_i|$}
      \State Let $\vec{w}$ be $j$-th axis of the space $[\vec{B}_i]$.
      \ForAll{$\vec{v} := \nabla\vec{F}_i[\vec{I}]/|\nabla\vec{F}_i[\vec{I}]|,
          \vec{B}_{i+1,1}, \ldots, \vec{B}_{i+1,|\vec{B}_{i+1}|}$}
          \label{alg:compute_next_basis:enum_vectors}
          \State $\vec{w} := \vec{w} - (\vec{w} \cdot \vec{v})\vec{v}$
          \label{alg:compute_next_basis:component_removal}
      \EndFor
      \If{$|w| \neq 0$}
        \label{alg:compute_next_basis:w}
        \State Append the vector $\vec{w}/|\vec{w}|$ to $\vec{B}_{i+1}$.
      \EndIf
    \EndFor
  \end{algorithmic}
\end{algorithm}


The last thing we must include to the computation of $\vec{B}_{i+1}$ is the
comparator $\vec{P}_i$. If $\vec{P}_i$ is $=$, then we are actually done. But in
all other cases we must extend $\vec{B}_{i+1}$ by the vector
$\nabla\vec{F}_i[\vec{I}]/|\nabla\vec{F}_i[\vec{I}]|$. We show the reason
for this extension in the following example.

\begin{example}
  \label{ex:local_space:constraint_motivation}
  Let us consider a trace $(x_1 \leq x_2, x_1 \geq 10)$. The coverage problem
  $\mathcal{S}=(\vec{F},\vec{P}, I)$ is thus the same as in
  Example~\ref{ex:local_space:1}, except that $\vec{P}=(\leq,\geq)$. The same is
  also the basis $\vec{B}_1$ and the gradient $\nabla\vec{F}_1[\vec{I}]$.
  However, the set of vectors $\vec{u} \in [\vec{B}_1]$ for which the predicate
  $\vec{P}_1(\vec{F}_1[\vec{I}+\vec{u}],0)$ evaluates to true is larger. We can
  see that in the Fig.\ref{fig:examples}(b), depicted as a half-plane (c.f. with
  Fig.\ref{fig:examples}(a)).

  Clearly, the basis $\vec{B}_2$ needs two basis vectors so that $[\vec{B}_2]$
  contains the whole half-plane. The basis vector $\vec{B}_{2,1}$ can be
  computed as we already described. The good choice for second basis vector is
  the (normalized) gradient vector, because it is then straightforward to
  express a \emph{constraint} for the half-space inside $[\vec{B}_2]$, i.e., all
  vectors $\vec{u} \in [\vec{B}_2]$ with $\vec{u}_2 \leq 0$.
\end{example}

\section{Constraints}
\label{sec:constraints}

Let $\mathcal{S}=(\vec{F},\vec{P}, I)$ be a coverage problem of the size
$n$. Example~\ref{ex:local_space:constraint_motivation} showed us that the
vector $\vec{B}$ of local spaces alone is not sufficient to accurately describe
the set of vectors keeping evaluation of predicates
$\vec{P}_i(\vec{F}_i[\vec{I}+\vec{u}^*],0)$ equal to
$\vec{P}_i(\vec{F}_i[\vec{I}],0)$, where $1 \leq i < n$ and $\vec{u} \in
[\vec{B}_n]$. We thus also need a vector $\vec{C}$ of dimension $n$, where each
$\vec{C}_i$ is a set of constraints allowing more accurate approximation of the
set of those vectors.

A \emph{constraint} is a triple $(\vec{n}, s, P)$, where $\vec{n}$ is a normal
vector, $s$ is a real number and $P$ is a comparator. It defines the set of all
vectors $\vec{u}$ for which the predicate $P(\vec{n} \cdot (\vec{u} -
s\vec{n}),0)$ is true. The predicate can be explained as follows. If the
comparator $P$ is $=$, then the predicate represents the equation $\vec{n} \cdot
(\vec{u} - s\vec{n}) = 0$, which is the well known formula for a hyper-plane.
Other comparators $P$ specify which of the two half-spaces separated by the
hyper-plane we consider. The hyper-plane itself may be included or excluded from
the considered half-space(s). 


We construct the vector $\vec{C}$ along with the vector $\vec{B}$. That is, once
$\vec{B}_{i+1}$ is computed (from $\vec{B}_i$, etc.) according to the algorithm
presented in the previous section, then we compute $C_{i+1}$ from $C_i,
\vec{P}_i, \vec{F}_i[\vec{I}], \nabla\vec{F}_i[\vec{I}]$, and $\vec{B}_{i+1}$ as
follows.

We start with $C_{i+1} = \emptyset$. If $\vec{P}_i$ is any comparator except
$=$, then we add the constraint $(\vec{e}_{|\vec{B}_{i+1}|},
-\frac{\vec{F}_i[\vec{I}]}{|\nabla\vec{F}_i[\vec{I}]|},\vec{P}_i)$ to
$\vec{C}_{i+1}$, where $\vec{e}_{|\vec{B}_{i+1}|} \in [\vec{B}_{i+1}]$ is the
$|\vec{B}_{i+1}|$-th axis vector. Observe, that $\vec{e}_{|\vec{B}_{i+1}|}$
corresponds to the vector $\vec{B}_{i,|\vec{B}_i|} =
\frac{1}{|\nabla\vec{F}_i[\vec{I}]|}\nabla\vec{F}_i[\vec{I}]$. It remain to
explain the computation of the number $s$.

Geometrically, $s$ defines the signed distance of the hyper-plane's point
$s\vec{n} \in [\vec{B}_{i+1}]$ from the origin, represented by the vector
$\vec{I}$. Note that $s\vec{n}$ corresponds to $\vec{I} + s\vec{n}^* \in
[\vec{B}_1]$. If we compute $s$ such that $\vec{F}_i[\vec{I} + s\vec{n}^*] = 0$,
then each vector $\vec{u}$ in some neighborhood of $s\vec{n}$ satisfying the
predicate $P(\vec{n} \cdot (\vec{u} - s\vec{n}),0)$ of the constraint will also
satisfy the corresponding predicate $P_i(\vec{F}_i[\vec{I}+\vec{u}^*],0)$ of the
coverage problem. Of course, the size of the neighborhood depends on the
function $\vec{F}_i$. If it is linear, then the neighborhood is the whole
hyper-plane. Otherwise, the neighborhood can be smaller; in the worst case it is
only $s\vec{n}$.


We can express $\vec{F}_i$ as a function $\vec{F}_i: [\vec{B}_{i+1}] \rightarrow
R$. We know that $(\vec{0}^\top, \vec{F}_i[\vec{I}])^\top \in \vec{F}_i$, where
$\vec{0} \in [\vec{B}_{i+1}]$ is the zero vector corresponding to the vector
$\vec{I}$ in the space $\vec{B}_1$ (because $\vec{I} + \vec{0}^* = \vec{I}$). We
also know the gradient $\nabla\vec{F}_i[\vec{I}] \in [\vec{B}_{i+1}]$. Since
$\vec{F}_i$ is not known to us, we linearly approximate it around
$(\vec{0}^\top, \vec{F}_i[\vec{I}])^\top$ using a hyper-plane $(\vec{0}^\top,
\vec{F}_i[\vec{I}])^\top + \sum_{j=1}^{|\vec{B}_{i+1}|}t_j(\vec{e}_j^\top,
\nabla_j\vec{F}_i[\vec{I}])^\top$, where each $\vec{e}_j \in [\vec{B}_{i+1}]$ is
the $j$-th axis vector and $t_j$ are any real numbers. There can be infinitely
many vectors $\vec{u} \in [\vec{B}_{i+1}]$ such that $(\vec{u}^\top,0)^\top$
lies in this hyper-plane. We search for one which lies on the line
$t(\nabla\vec{F}_i[\vec{I}]^\top, 0)^\top$, where $t \in R$. So, we solve the
equation 
$$
  t\begin{pmatrix} \nabla\vec{F}_i[\vec{I}] \\ 0 \end{pmatrix}
  = 
  \begin{pmatrix} \vec{0} \\ \vec{F}_i[\vec{I}] \end{pmatrix}
  + \sum_{j=1}^{|\vec{B}_{i+1}|}
  t_j\begin{pmatrix} \vec{e}_j \\ \nabla_j\vec{F}_i[\vec{I}] \end{pmatrix}
$$
It can be decomposed into a system of two equations
$$
  t\nabla\vec{F}_i[\vec{I}]^\top = (t_1,\ldots,t_{|\vec{B}_{i+1}|})^\top,~~~~~~~~~~~~~
  0 = \vec{F}_i[\vec{I}] + \sum_{j=1}^{|\vec{B}_{i+1}|}t_j\nabla_j\vec{F}_i[\vec{I}]
$$
The first equation can further be decomposed into $|\vec{B}_{i+1}|$ equations
$$
  t\nabla_1\vec{F}_i[\vec{I}] = t_1,~\ldots,~
  t\nabla_{|\vec{B}_{i+1}|}\vec{F}_i[\vec{I}] = t_{|\vec{B}_{i+1}|}
$$
If we substitute all these equations to the equation above (we substitute
numbers $t_j$), then we get a single equation of one unknown $t$:
$$
  0 = \vec{F}_i[\vec{I}] + \sum_{j=1}^{|\vec{B}_{i+1}|}(t\nabla_j\vec{F}_i[\vec{I}])\nabla_j\vec{F}_i[\vec{I}]
$$
We solve for $t$ as follows
$$
  \begin{array}{rl}
    0 =& \vec{F}_i[\vec{I}] + t\sum_{j=1}^{|\vec{B}_{i+1}|}(\nabla_j\vec{F}_i[\vec{I}])^2\\
    0 =& \vec{F}_i[\vec{I}] + t|\nabla\vec{F}_i[\vec{I}]|^2\\
    t =& -\vec{F}_i[\vec{I}] / |\nabla\vec{F}_i[\vec{I}]|^2
  \end{array}
$$
Lastly, recall that the normal of the constraint $\vec{e}_{|\vec{B}_{i+1}|} \in
[\vec{B}_{i+1}]$ corresponds to vector $\vec{B}_{i,|\vec{B}_i|} =
\frac{1}{|\nabla\vec{F}_i[\vec{I}]|}\nabla\vec{F}_i[\vec{I}] \in [\vec{B}_i]$.
That means, $\nabla\vec{F}_i[\vec{I}] =
|\nabla\vec{F}_i[\vec{I}]|\vec{B}_{i,|\vec{B}_i|}$. Therefore, the line
$t(\nabla\vec{F}_i[\vec{I}]^\top, 0)^\top$ we started with can equivalently be
expressed as $t|\nabla\vec{F}_i[\vec{I}]|(\vec{B}_{i,|\vec{B}_i|}, 0)^\top$. So,
we get $s=t|\nabla\vec{F}_i[\vec{I}]| =
-\vec{F}_i[\vec{I}]/|\nabla\vec{F}_i[\vec{I}]|$.

\begin{example}
  \label{ex:constraints:constraints_set}
  Let us consider a trace $(x_1 \leq x_2, x_1 = 1)$. Then we have
  $\vec{F}=([x_1-x_2],[x_1-1])$, $\vec{P}=(\leq,=)$ and with a valuation $I$
  mapping $x_1$ to 0 and $x_2$ to 1 we get a coverage problem
  $\mathcal{S}=(\vec{F},\vec{P}, I)$ of size 2. We further have
  $\vec{F}_1[\vec{I}] = -1$, the gradient $\nabla\vec{F}_1[\vec{I}] =
  (1,-1)^\top$ and the basis $\vec{B}_2=\{ \vec{B}_{2,1} =
  \frac{1}{\sqrt{2}}(1,1)^\top, \vec{B}_{2,2} = \frac{1}{\sqrt{2}}(1,-1)^\top
  \}$. The set $\vec{C}_1$ is (always) empty. The set $\vec{C}_2 = \{
  (\vec{n}=(0,1)^\top, s=\frac{1}{\sqrt{2}}, P=\leq) \}$. The constraint is
  graphically depicted in Fig.\ref{fig:examples}(c).
\end{example}

The next step in the construction of $\vec{C}_{i+1}$ is to transform all
constraints in $\vec{C}_i$ to the space $[\vec{B}_{i+1}]$ and insert them to
$\vec{C}_{i+1}$. Let $(\vec{n}, s, P) \in \vec{C}_i$.

First we transform the vector $\vec{n} \in [\vec{B}_i]$ to a vector $\vec{m} \in
[\vec{B}_{i+1}]$. Since the basis $\vec{B}_{i+1}$ is orthonormal, the
transformation is straightforward; for each $1 \leq j \leq |\vec{B}_{i+1}|$ the
$j$-th coordinate of $\vec{m}$ is $\vec{m}_j = \vec{n} \cdot \vec{B}_{i+1,j}$.
Note that if we transform the vector $\vec{m}$ back to the space $[\vec{B}_i]$,
then we get the vector $\vec{n}' =
\sum_{j=1}^{|\vec{B}_{i+1}|}\vec{m}_j\vec{B}_{i+1,j}$, which can differ from the
original normal vector $\vec{n}$ (if $|\vec{B}_{i+1}| < |\vec{B}_i|$).

Next we must also update the distance $s$, because for the transformed normal
$\vec{m}$ we can have $\vec{m}^* \neq \vec{n}^*$. We compute new value, say $r$,
as the signed distance from the origin to the intersection point of the
hyper-plane $\vec{n} \cdot (\vec{u} - s\vec{n}) = 0$ and the line $\vec{u} = r
\vec{n}'$. Observe, that both equations are expressed in the space
$[\vec{B}_i]$. If we substitute the line equation to the equation of the
hyper-plane (we substitute the vector $\vec{u}$), then we get the equation
$\vec{n} \cdot (r \vec{n}' - s\vec{n}) = 0$. If $\vec{n} \cdot \vec{n}' \neq 0$,
then it has a solution $r = s\frac{\vec{n} \cdot \vec{n}}{\vec{n} \cdot
\vec{n}'}$. Otherwise, we cancel our attempt to (transform and) insert the
constraint to $\vec{C}_{i+1}$. Of course, this may lead to a reduction of
performance boost we expect from the resulting vector $\vec{C}$. Finally, if
$\vec{m}, r$ are successfully computed, then we insert the constraint $(\vec{m},
r, P)$ to $\vec{C}_{i+1}$.

\begin{example}
  \label{ex:constraints:constraints_set2}
  Let us extend the trace in Example~\ref{ex:constraints:constraints_set} by
  appending one \abe $x_2 = x_1 + 3$ to it. So, we get the trace $(x_1 \leq x_2,
  x_1 = 1, x_2 = x_1 + 3)$. We must also change the valuation $I$ for the
  coverage problem, for example, such that both $x_1,x_2$ are mapped to 1. Then
  the base $\vec{B}_2$ is the same as in
  Example~\ref{ex:constraints:constraints_set}. However, $\vec{C}_2 = \{
  (\vec{n}=(0,1)^\top, s=0, P=\leq) \}$. Further, $\nabla\vec{F}_2[\vec{I}] =
  \frac{1}{\sqrt{2}}(1,1)^\top$, $\vec{B}_3 = \{ \vec{B}_{3,1} =
  \frac{1}{\sqrt{2}}(1,-1)^\top \}$ and $\vec{B}_{3,1}^* = (0,1)^\top$. Our goal
  is to compute the set $\vec{C}_3$. That involves the transformation of the
  constraint in $\vec{C}_2$ to the space $[\vec{B}_3]$. The vector $\vec{m} =
  (\vec{n} \cdot \vec{B}_{3,1})^\top = ((0,1)^\top \cdot
  \frac{1}{\sqrt{2}}(1,-1)^\top)^\top = (-\frac{1}{\sqrt{2}})^\top$. Observe,
  that $\vec{n}' = -\frac{1}{\sqrt{2}}\frac{1}{\sqrt{2}}(1,-1)^\top =
  \frac{1}{2}(-1,1)$. Since $s = 0$, then also $r = 0$. Therefore, we get
  $((-\frac{1}{\sqrt{2}})^\top,0,\leq) \in \vec{C}_3$. The constraint is
  depicted in the Fig.\ref{fig:examples}(d).
\end{example}

We say that a vector $\vec{u} \in [\vec{B}_i]$ satisfies a constraint $(\vec{n},
s, P) \in \vec{C}_i$, if the predicate $P(\vec{n} \cdot (\vec{u} - s\vec{n}),0)$
is true. And $\vec{u}$ satisfies constraints $\vec{C}_i$, if either $\vec{C}_i =
\emptyset$ or it satisfies all constraints in $\vec{C}_i$.


\subsection{Clipping}
\label{sec:clipping}

Let $\vec{u} \in [\vec{B}_i]$ be a vector and $(\vec{n}, s, P) \in \vec{C}_i$ be
a constraint. If $\vec{u}$ satisfies the constraint, then the clipped $\vec{u}$
is the vector $\vec{u}$ itself. Otherwise, the clipped $\vec{u}$ is a projection
of $\vec{u}$ in the direction $\vec{n}$ to a vector satisfying the constraint.
Namely, it is the vector $\vec{u} + (s - \frac{\vec{u} \cdot \vec{n}}{\vec{n}
\cdot \vec{n}} + h(P))\vec{n}$, where the function $h = \{ (\neq,\epsilon),
(<,-\epsilon), (\leq,0), (>,\epsilon), (\geq,0) \}$. The number $\epsilon$ is
computed as presented in Section~\ref{sec:epsilon_step}. It represents the
smallest step from the number $\frac{\vec{u} \cdot \vec{n}}{\vec{n} \cdot
\vec{n}}$. We can easily check that vector $\vec{v} = \vec{u} + (s -
\frac{\vec{u} \cdot \vec{n}}{\vec{n} \cdot \vec{n}})\vec{n}$ lies in the
hyper-plane of the constraint, i.e. $\vec{n} \cdot (\vec{v} - s\vec{n}) = 0$.
The function $h$ thus provides a shift from the hyper-plane to the desired
half-space according to $P$.

Clipping of a vector $\vec{u}$ by constraints in $\vec{C}_i$ proceeds in
iterations. In each iteration $\vec{u}$ is clipped by each constraint in
$\vec{C}_i$. The algorithm terminates either if $\vec{u}$ satisfies all
constraints in $\vec{C}_i$ or the number of iterations exceeds a given limit. In
our implementation the limit is 10. The limit is necessary, because if the set
of vectors satisfying all constraints is empty or concave, then the process
would diverge or converge slowly (may even require infinite number of
iterations).

\begin{remark}
  The vector $\vec{u}$ we clip may be a candidate to a solution of the coverage
  problem. Its projection along constraint's normal vector $\vec{n}$ may change
  the value $\vec{F}_i[\vec{I}+\vec{u}^*]$, if $\nabla \vec{F}_i[\vec{I}]$ is
  not orthogonal to $\vec{n}$. It may thus be more effective to project
  $\vec{u}$ along the component of $\vec{n}$ orthogonal to $\nabla
  \vec{F}_i[\vec{I}]$. We can easily compute the component as $\vec{m} = \vec{n}
  - \frac{\vec{n} \cdot \nabla \vec{F}_i[\vec{I}]}{\nabla \vec{F}_i[\vec{I}]
  \cdot \nabla \vec{F}_i[\vec{I}]}\nabla \vec{F}_i[\vec{I}]$. The projected
  $\vec{u}$ is then the vector $\vec{u} + (s\frac{\vec{n} \cdot \vec{n}}{\vec{n}
  \cdot \vec{m}} - \frac{\vec{u} \cdot \vec{n}}{\vec{n} \cdot \vec{m}} +
  h(P))\vec{m}$. Unfortunately, this optimization has strongly negative impact
  on the convergence of the algorithm, and so we apply this projection only in
  the first iteration.
\end{remark}

\section{Solving coverage problem}
\label{sec:searching}

Let $\mathcal{S}=(\vec{F},\vec{P}, I)$ be a coverage problem with a size $n$. In
this section we present an algorithm searching for a solution of $\mathcal{S}$
effectively.

The algorithm is iterative. It builds a finite sequence of valuations $I_1,
\ldots, I_m$. The goal of an iteration $i$ is to compute the valuation
$I_{i+1}$. The valuation $I_1$ is the initial valuation $I$ and each valuation
$I_i$, where $1 \leq i < m$, is such that $\mathcal{S}_i = (\vec{F},\vec{P},
I_i)$ is a coverage problem with the size $n$. If the valuation $I_m$ is a
solution of $\mathcal{S}$, then the algorithm terminates with success.
Otherwise, it terminated with failure. One possible failure is that the
algorithm exhausted all available resources, like time for the analysis or
maximal length of the constructed sequence of valuations. Another possible
failure is that the algorithm failed to produce a valuation $I_{m}$ which is
either a solution of $\mathcal{S}$ or forming (together with vectors
$\vec{F},\vec{P}$) a coverage problem $(\vec{F},\vec{P}, I_m)$.

It remains to discuss the algorithm of one iteration, i.e., how to compute a
valuation $I_{i+1}$ from the coverage problem $\mathcal{S}_i = (\vec{F},\vec{P},
I_i)$. The algorithm first computes the vector $\vec{B}$ of orthonormal bases
(according to Section~\ref{sec:local_spaces}) and the vector $\vec{C}$ of sets
of constraints (according to Section~\ref{sec:constraints}). Besides boosting
the search, these two vectors also simplify the computation, because we can
consider only their last coordinates, i.e., $\vec{B}_n$ and $\vec{C}_n$ and also
only the last coordinates $\vec{F}_n$ and $\vec{P}_n$ of vectors $\vec{F}$ and
$\vec{P}$. For that reason we denote $\vec{B}_n, \vec{C}_n, \vec{F}_n,
\vec{P}_n$ as $B, C, F, P$, respectively.

Then we apply three algorithms, namely \emph{gradient descent step},
\emph{bit-mutations}, and \emph{random values}, executed sequentially in that
order. Each algorithm produces a sequence of vectors from $[B]$. We concatenate
these sequences into one sequence $\vec{u}_1, \ldots, \vec{u}_k \in [B]$ and we
process these vectors sequentially one by one. For a vector $\vec{u}_j$, where
$1 \leq j \leq k$, we call the functions $\vec{F}[\vec{I}_i + \vec{u}_j^*]$.
There are two possible outcomes; either we obtain the value $F[\vec{I}_i +
\vec{u}_j^*]$ or not. In the former case, if $P(F[\vec{I}_i + \vec{u}_j^*], 0)$
is true, then $\vec{I}_i + \vec{u}_j^*$ is a solution of the coverage problem
$\mathcal{S}$; we set $\vec{I}_{i+1}$ as $\vec{I}_i + \vec{u}_j^*$ and terminate
the entire algorithm with success. Otherwise, we check whether $F[\vec{I}_i +
\vec{u}_j^*]$ is closer to satisfying the predicate above than the original
value $F[\vec{I}_i]$. The check depends on the comparator $P$ and it is defined
in the table below ($P$ is in the left column).
\begin{center}
  \begin{tabular}{rl}
    $=$ ~~~~&~~~~ $|F[\vec{I}_i + \vec{u}_j^*]| < |F[\vec{I}_i]|$ \\
    $\neq$ ~~~~&~~~~ $|F[\vec{I}_i + \vec{u}_j^*]| > |F[\vec{I}_i]|$ \\
    $<,\leq$ ~~~~&~~~~ $F[\vec{I}_i + \vec{u}_j^*] < F[\vec{I}_i]$ \\
    $>,\geq$ ~~~~&~~~~ $F[\vec{I}_i + \vec{u}_j^*] > F[\vec{I}_i]$ \\
  \end{tabular}
\end{center}
If the check passes, then we set $\vec{I}_{i+1}$ as $\vec{I}_i + \vec{u}_j^*$
and we are done with the iteration. Otherwise, we continue with processing of
the next vector $\vec{u}_{j+1}$, if $j < k$. Otherwise, whole algorithm
terminates with failure, because we failed to compute the valuation
$\vec{I}_{i+1}$. Lastly, when we fail to obtain the value $F[\vec{I}_i +
\vec{u}_j^*]$, then we proceed with the next vector $\vec{u}_{j+1}$, if $j < k$.
Otherwise, whole algorithm terminates with failure.

\begin{remark}
  Before we discuss details of the individual algorithms in the following
  sub-sections, we want to stress that they are basically re-implementations of
  corresponding algorithms in the original version of
  \fizzer~\cite{fizzer_testcomp_2024} in a new setting, i.e., to work in the
  local space $[B]$ while being aware of the constraints in $C$. Originally,
  Bit-mutations were primarily used for detection of sensitive bytes. Here they
  are just candidates for a solution. The typed gradient descent algorithm in
  the original version is represented here by the Random values algorithm,
  providing the generation of random seeds, and the Gradient descent step
  algorithm, providing one gradient descent step.
\end{remark}

\subsection{Gradient descent step}
\label{sec:grad_step}

Our goal is to find a vector $p\nabla F[\vec{I}]$, where $p$ is a real number,
such that $P(F[\vec{I} + p\nabla F[\vec{I}]^*], 0)$ is true. Recall that we
solved similar task in Section~\ref{sec:constraints} (computation of
constraint's parameter). There we computed a real number $t$ satisfying
$F[\vec{I} + t\nabla F[\vec{I}]^*] = 0$ (under the assumption of linearity of
$F$) and we obtained the solution $t = -F[\vec{I}] / |\nabla F[\vec{I}]|^2$.
With $t$ available we can compute possible values for $p$ according to the
following mapping from the predicate $P$ to values for $p$.
$$
  = \rightarrow t~~~~
  \neq \rightarrow t-\epsilon, t+\epsilon~~~~
  < \rightarrow t-\epsilon~~~~
  \leq \rightarrow t-\epsilon, t~~~~
  > \rightarrow t+\epsilon~~~~
  \geq \rightarrow t, t+\epsilon
$$
where the number $\epsilon$ s computed according to algorithm in
Section~\ref{sec:line_epsilon_step}. This algorithm is started with an initial
value for $\epsilon$, computed by algorithm in Section~\ref{sec:epsilon_step},
from the value $z = (1-\alpha) \max \{ |\vec{I}_j + t\nabla
F[\vec{I}]_j^*|~;~1\leq j \leq \nu(\vec{F}) \} + \alpha |F[\vec{I}]^*]|$, where
$\alpha = 0.01$ in our implementation. The number $z$ is a linear interpolant of
two numbers, because success or failure of the computation of $F[\vec{I} +
p\nabla F[\vec{I}]^*]$ may depend on their magnitudes.

Depending on the predicate we can have either one or two values for the number
$p$. Therefore, the algorithm produces at most two vectors. However, before they
are returned, each of them is clipped by constraints in $C$, as explained in
Section~\ref{sec:clipping}.

\begin{remark}
  In our implementation, beside the step in the direction $\nabla F[\vec{I}]$ we
  also perform steps in each direction $\nabla_j F[\vec{I}] \vec{e}_j$, where $1
  \leq j \leq |B|$ and $\vec{e}_j$ is the $j$-th axis vector. Note that for each
  $j$ we get $t_j = -F[\vec{I}] / |\nabla_j F[\vec{I}]\vec{e}_j|^2$. This
  extension is for the consistency with the original version of \fizzer. It
  improves the ability of the gradient descent to escape from local minima of
  $F$. However, in the worst case the algorithm can produce $2(1+|B|)$ vectors.
\end{remark}

\subsection{Bit-mutations}
\label{sec:bit_mutations}

The goal of this algorithm to compute one output vector for each bit of each
parameter (input variable) of the black-box function $F$. Clearly, the algorithm
produces at most $64|\nu(F)|$ vectors.

It is sufficient to describe how to compute the vector for one variable $x_i \in
\nu(F)$ and the $j$-th least significant bit of $x_i$, denoted as $x_{i,j}$. We
ignore the type $\tau(x_i)$. More precisely, we trait the value in $x_i$ as if
it was of the unsigned integer type with the same number of bits as in
$\tau(x_i)$.

Let $y = (1 - 2x_{i,j}) 2^{j-1}$. We can easily check that $x_i + y$ is the
value $x_i$ with the bit $x_{i,j}$ flipped. All vectors $\vec{w} \in
[\vec{B}_1]$ where $x_i$ has the bit $x_{i,j}$ flipped are solutions of the
plane equation $\vec{B}_{1,i} \cdot (\vec{w} - y\vec{B}_{1,i}) = 0$. Our goal is
to find a vector $\vec{u} \in [B]$ such that $\vec{B}_{1,i} \cdot (\vec{u}^* -
y\vec{B}_{1,i}) = 0$ and for each vector $\vec{v} \in [B]$ such that
$\vec{B}_{1,i} \cdot (\vec{v}^* - y\vec{B}_{1,i}) = 0$ we have $|\vec{u}^* -
y\vec{B}_{1,i}| \leq |\vec{v}^* - y\vec{B}_{1,i}|$.

\begin{example}
  We provide intuition behind the definition of the vector $\vec{u}$ above on
  two pictures in Fig.\ref{fig:examples}(e) and (f). In both pictures the plane
  $\vec{B}_{1,i} \cdot (\vec{w} - y\vec{B}_{1,i}) = 0$ is blue, the space $[B]$
  (transformed to $[\vec{B}_1]$) is orange, the set of all possible solutions is
  the green line and the solution $\vec{u}$ we search for is the green dot. In
  the figure (e) we have a case $|\vec{B}_1|=2, |B|=1$. There is only one
  solution. In the figure (f) we have a case $|\vec{B}_1|=3, |B|=2$. There are
  infinitely many solutions. The one we search for is the closest solution to
  $y\vec{B}_{1,i}$.
\end{example}

We search for the solution vector $\vec{u} \in [B]$ such that we apply gradient
descent of a function $f(\vec{u}) = |\vec{u}^* - y\vec{B}_{1,i}|^2$ to find its
minimum. We must also keep the vector $\vec{u}^*$ inside the plane, i.e.,
$\vec{B}_{1,i} \cdot (\vec{u}^* - y\vec{B}_{1,i}) = 0$. Therefore, after each
gradient descent step, which may move $\vec{u}^*$ out of the plane, we
immediately move $\vec{u}^*$ back to the plane. We start the descent from
$\vec{u} = y\vec{e}_p$, where $\vec{e}_p$ is the $p$-th axis vector and $p$ is
the index any basis vector $B_p \in B$ such that $B_p^* \cdot \vec{B}_{1,i} \neq
0$. If there is no such basis vector in $B$, then there is no solution and we do
not generate a vector for that bit. Otherwise, as the gradient descent proceeds
the vector $\vec{u}$ converges towards the solution. In our implementation we
always apply 10 gradient descent steps. However, smarter termination condition
can be established.

It remains to discuss three things. First, how to move the vector $\vec{u}$ back
to the plane after each gradient descent step. Next, we must compute partial
derivatives of the function $f$, and lastly we must compute the parameter $t$
for the gradient step vector $t \nabla f$. Before we start we must realize that
the vector $\vec{u}$ can be expressed as a linear combination of basis vectors,
i.e., as $\vec{u} = \sum_{k=1}^{|B|}\vec{u}_k B_k$.

Let us rewrite the plane equation as follows.
$$
  \begin{array}{rl}
    \vec{B}_{1,i} \cdot (\vec{u}^* - y\vec{B}_{1,i}) &= 0 \\
    \vec{B}_{1,i} \cdot \vec{u}^* &= y \\
    \vec{B}_{1,i} \cdot (\sum_{k=1}^{|B|}\vec{u}_k B_k)^* &= y \\
    \vec{B}_{1,i} \cdot \sum_{k=1}^{|B|}\vec{u}_k B_k^* &= y \\
    \sum_{k=1}^{|B|}\vec{u}_k (B_k^* \cdot \vec{B}_{1,i}) &= y \\
    \vec{u}_p (B_p^* \cdot \vec{B}_{1,i}) &= y - 
      \sum_{k=1}^{p-1}\vec{u}_k (B_k^* \cdot \vec{B}_{1,i}) -
      \sum_{k=p+1}^{|B|}\vec{u}_k (B_k^* \cdot \vec{B}_{1,i}) \\
    \vec{u}_p &= y - 
      \sum_{k=1}^{p-1}\vec{u}_k \frac{B_k^* \cdot \vec{B}_{1,i}}{B_p^* \cdot \vec{B}_{1,i}} -
      \sum_{k=p+1}^{|B|}\vec{u}_k \frac{B_k^* \cdot \vec{B}_{1,i}}{B_p^* \cdot \vec{B}_{1,i}} \\
    \vec{u}_p &= y - 
      \sum_{k=1}^{p-1}\vec{u}_k \frac{B_{k,i}^*}{B_{p,i}^*} -
      \sum_{k=p+1}^{|B|}\vec{u}_k \frac{B_{k,i}^*}{B_{p,i}^*} \\
  \end{array}
$$
To move $\vec{u}$ back to the plane we just recompute its coordinate $\vec{u}_p$
according to the last equation above.

Next, we simplify the function $f$ as follows.
$$
  \begin{array}{rl}
    f(\vec{u})
      &= |\vec{u}^* - y\vec{B}_{1,i}|^2
       = |\sum_{k=1}^{|B|}\vec{u}_k B_k^* - y\vec{B}_{1,i}|^2 \\
      &= \sum_{k=1}^{|B|}(\sum_{l=1}^{|B|} \vec{u}_k \vec{u}_l (B_k^* \cdot B_l^*)
                          - \vec{u}_k y (B_k^* \cdot \vec{B}_{1,i}))
         - \sum_{k=1}^{|B|}\vec{u}_k y (B_k^* \cdot \vec{B}_{1,i})
         + y^2 \\
      &= \sum_{k=1}^{|B|}\sum_{l=1}^{|B|}\vec{u}_k \vec{u}_l (B_k^* \cdot B_l^*)
         - \sum_{k=1}^{|B|}\vec{u}_k (1+y) (B_k^* \cdot \vec{B}_{1,i})
         + y^2 \\
      &= \sum_{k=1}^{|B|}\vec{u}_k^2
         - (1+y)\sum_{k=1}^{|B|}\vec{u}_k (B_k^* \cdot \vec{B}_{1,i})
         + y^2 \\
      &= \sum_{k=1}^{|B|}\vec{u}_k^2
         - (1+y)\sum_{k=1}^{|B|}\vec{u}_k B_{k,i}^*
         + y^2 \\
  \end{array}
$$
Now we can compute partial derivatives $\nabla_k f(\vec{u})$ for all $1 \leq k
\leq |B|$ such that $k \neq p$. That is because $\vec{u}_p$ is always recomputed
as shown above. Therefore, we can keep $\nabla_p f(\vec{u}) = 0$.
$$
  \begin{array}{rl}
    \nabla_k f(\vec{u})
      &= \sum_{l=1}^{|B|}2\vec{u}_l \frac{\partial \vec{u}_l}{\partial \vec{u}_k}
         - (1+y)\sum_{l=1}^{|B|}B_{k,i}^* \frac{\partial \vec{u}_l}{\partial \vec{u}_k} \\
  \end{array}
$$
Clearly, for each $l \neq p$ we have $\frac{\partial \vec{u}_l}{\partial
\vec{u}_k}$ is 1 if $l = k$ and 0 otherwise. The partial $\frac{\partial
\vec{u}_p}{\partial \vec{u}_k}$ is computed as follows.
$$
  \frac{\partial \vec{u}_p}{\partial \vec{u}_k} =
    - \sum_{l=1}^{p-1}\frac{B_{k,i}^*}{B_{p,i}^*}\frac{\partial \vec{u}_l}{\partial \vec{u}_k}
    - \sum_{l=p+1}^{|B|}\frac{B_{k,i}^*}{B_{p,i}^*}\frac{\partial \vec{u}_l}{\partial \vec{u}_k}
  = -\frac{B_{k,i}^*}{B_{p,i}^*}
$$
If we substitute the computed partials of $\vec{u}$'s coordinates to $\nabla_k
f(\vec{u})$, then we get the following.
$$
  \nabla_k f(\vec{u})
    = 2\vec{u}_k + 2\vec{u}_p(-\frac{B_{k,i}^*}{B_{p,i}^*})
       - (1+y)(B_{p,i}^*(-\frac{B_{k,i}^*}{B_{p,i}^*}) + B_{k,i}^*)
    = 2(\vec{u}_k - \vec{u}_p\frac{B_{k,i}^*}{B_{p,i}^*})
$$

Lastly, we need to compute a parameter $t$ for the gradient vector $t\nabla
f(\vec{u})$. In our implementation we proceed similarly to computation of
constraint's parameter in Section~\ref{sec:constraints}. Here we also linearly
approximate the function $f$\footnote{The function $f$ is known, so smarter
approach can be used.} at the point $(\vec{u}^\top, f(\vec{u}))^\top \in f$ via
a hyper-plane $(\vec{u}^\top, f(\vec{u})^\top) +
\sum_{k=1}^{|B|}t_k(\vec{e}_k^\top, \nabla f(\vec{u}))^\top$. And similarly we
compute its intersection with the line $(\vec{u}^\top, 0)^\top + t(\nabla
f(\vec{u})^\top, 0)^\top$. We thus end up with a solution in the same form,
i.e., $t = -f(\vec{u}) / |\nabla f(\vec{u})|^2$.

\begin{remark}
  Each computed vector should be clipped by constraints in $C$, as described in
  Section~\ref{sec:clipping}. However, we do not do so in our implementation,
  because we empirically discovered that not clipped valuations (more vulnerable
  to early escaping from the path) have a reasonable potential to accidentally
  solve other coverage problems (for other execution paths).
\end{remark}

\subsection{Random values}
\label{sec:random_values}

Each vector $\vec{u} \in [B]$ can be expressed as $\vec{u} =
\sum_{i=1}^{|B|}t_iB_i$, for some real numbers $t_i$. Random generation of
vectors in $[B]$ thus reduces to random generation of the numbers $t_i$.
However, the sample space of all these numbers together is huge. We should
rather sample specific areas in $[B]$.

There are two areas in $[B]$ which may possibly comprise a solution to the
coverage problem. First of all, it is the area around the current vector
$\vec{I} \in [\vec{B}_1]$, because that is where the entire algorithm converged
to so far. This vector corresponds to the origin of $[B]$, which is the zero
vector $\vec{0}$.

The second area worth sampling is around the location computed by the gradient
descent step algorithm, presented in Section~\ref{sec:grad_step}, because the
algorithm in fact predicts a possible solution, by linear approximation of $F$.
The location is the vector $-\frac{F[\vec{I}]}{|\nabla F[\vec{I}]|^2}\nabla
F[\vec{I}]$, which we denote as $\vec{g}$ here.

We represent a random sampled area by a hyper-cube $\vec{u} =
\sum_{i=1}^{|B|}t_iB_i + \vec{d}$, where the vector $\vec{d} \in [B]$ is the
center of the cube. The vector $\vec{d}$ can thus either be $\vec{0}$ or
$\vec{g}$. Let $a$ be a positive real number defining the half size of cube's
edge. Then we can uniformly sample numbers $t_i$ from the range
$\langle-a,a\rangle$.

A reasonable value for the number $a$ can be inferred from $|F[\vec{I}]|$ under
assumption the function $|F|$ decays quickly from the current value
$|F[\vec{I}]|$ to zero with the distance from the cube's center. So, the number
$a$ should be a logarithm of $|F[\vec{I}]|$. In our implementation we use $a =
100\ln(|F[\vec{I}]| + 1)$.

Each vector obtained from random sampling each cube should also be clipped by
constraints in $C$, as described in Section~\ref{sec:clipping}.

\begin{remark}
  Similarly to vectors generated by Bit-mutations algorithm, vectors generated
  here also have a potential to accidentally solve other coverage problems, if
  they are not clipped. For that reason our implementation produces each
  generated vector in two versions; clipped and not clipped. Lastly, we sample
  each hyper-cube 100 times. The algorithm thus produces 400 vectors.
\end{remark}

\section{Implementation}
\label{sec:implementation}

The algorithm presented in previous sections works with real numbers. However,
the input variables $\nu(\vec{F})$ and also variables in the actual
implementation of the algorithm hold numbers with a infinite precision. This
must be taken into account. In two subsections below we define algorithms
providing robust computation of an $\epsilon$-step from given value or along a
line. The algorithms are important for numerical stability of computations
discussed in sections above.

\begin{example}
  Let $x$ be a variable of 64-bit floating point type. We want to compute a
  value $x + \epsilon$ which is close to $x$, but we also want $x \neq x +
  \epsilon$. If the number $\epsilon$ is a constant, say $\epsilon = 1$, then
  for large values of $x$, like $x = 2^{60} (\approx 10^{18})$, we have $x = x +
  \epsilon$. Therefore, $\epsilon$ must be a function of $x$.
\end{example}

\subsection{$\epsilon$-step from value}
\label{sec:epsilon_step}

Given a 64-bit floating point value $a$ our goal is to compute a 64-bit floating
point value $\epsilon$ such that $a+\epsilon$ differs from $a$ in the lower half
of all representable digits. Let $d$ be the number of all representable digits.
Since $a$ can always be expressed in form $m2^n$, where $m$ is a floating point
number satisfying $-1 \leq m < 0.5 \vee 0.5 < m \leq 1$ and $n$ is an integer
number, we can compute $\epsilon$ as the number $2^{n - \lfloor d/2 \rfloor}$.
Note that $\epsilon$ computed this way is considerably smaller than $a$ (by
$\lfloor d/2 \rfloor$ orders of magnitude with the base 2), yet it also modifies
reasonable amount of digits for affecting results of further computations.

\subsection{$\epsilon$-step along line}
\label{sec:line_epsilon_step}

Let us start with a notation. Given a vector $\vec{J} \in [\vec{B}_1]$, we
denote $\langle \vec{J} \rangle \in [\vec{B}_1]$ a vector such that its $i$-th
coordinate $\langle \vec{J} \rangle_i$, where $1 \leq i \leq |\vec{B}_1|$, is
computed from $\vec{J}_i$ such that it is converted to a value of the type
$\tau(x_i)$ (see Section~\ref{sec:embedding_valuations} for details) and that
value is further converted back to a real number.

Let $\vec{I}, \vec{g} \in [\vec{B}_1]$ be vectors. Our goal is to compute a
64-bit floating point value $\epsilon$ such that $\langle \vec{I} +
\epsilon\vec{g}\rangle \neq \langle \vec{I} \rangle$ and the maximum of two
distances described next is minimal. One distance is $|\epsilon\vec{g}|$. The
other is the distance of $\langle \vec{I} + \epsilon\vec{g}\rangle$ from the
line $\vec{I} + t\vec{g}$, where $t$ is any real number.

Finding an exact solution is difficult. So, we rather compute a vector $\vec{S}$
of $m$ sample points on the line ($m = 2|\vec{B}_1|$ in our implementation),
where $\vec{S}_i = \vec{I} + \epsilon_i\vec{g}$, and we return epsilon of the
sample with the minimal maximum of the distances.

Since the algorithm accepts, besides $\vec{I}$ and $\vec{g}$, also the value
$\epsilon_1$, the sample $\vec{S}_1$ can be computed directly. Then, from the
known sample $\vec{S}_i$ we compute $\vec{S}_{i+1} = \vec{S}_i +
\min\{\frac{\chi(\langle\vec{S}_i\rangle_j,
\textrm{sign}(\vec{d}_j))}{\vec{g}_j}~;~1\leq j \leq |\vec{B}_1|\} \vec{g}$,
where the function $\chi(a,s)$ returns the next value after $a$ in the ordering
of the type $\tau(x_j)$ in the direction given by the sign $s$. In our
implementation we use the function \texttt{nextafter} from the standard
\texttt{C} library. With $\vec{S}_{i+1}$ computed, we can easily compute also
$\epsilon_{i+1} = \frac{(\vec{S}_{i+1} - \vec{I}) \cdot \vec{g}}{\vec{g} \cdot
\vec{g}}$. The computation of $|\epsilon_{i+1}\vec{g}|$ is straightforward. We
can compute the distance of $\langle\vec{S}_{i+1}\rangle$ to the line as the
distance from $\langle\vec{S}_{i+1}\rangle$ to the intersection point $\vec{u}$
of the line $\vec{u} = \vec{I} + t\vec{g}$ and the plane $\vec{g} \cdot (\vec{u}
- \langle\vec{S}_{i+1}\rangle) = 0$. By solving these equations we get $t =
\frac{(\langle\vec{S}_{i+1}\rangle - \vec{I}) \cdot \vec{g}}{\vec{g} \cdot
\vec{g}}$.

\begin{remark}
  The entire algorithm is implemented in C++. The source code is available in
  \fizzer's Git repository~\cite{fizzer_repo} under tag \texttt{TESTCOMP25}. The
  source code of the original version of the tool is under tag
  \texttt{TESTCOMP24}.
\end{remark}

\section{Evaluation}
\label{sec:evaluation}

\addtolength{\abovecaptionskip}{-12mm}
\begin{wrapfigure}[21]{r}{0pt}
  \raisebox{10mm}[\height]{\includegraphics[width=5.6cm]{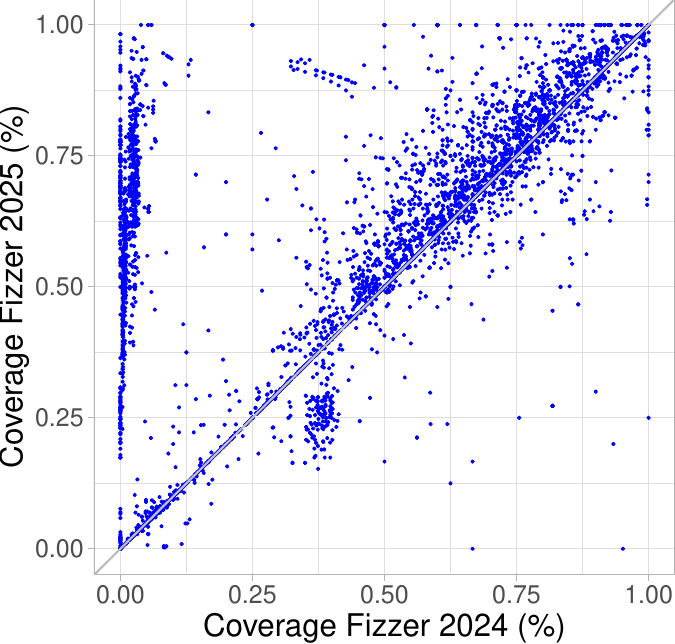}}
  \caption{Comparison of branch coverages achieved by the original \fizzer
    (horizontal axis) and the new one (vertical axis). Each dot represents one
    benchmark. For benchmarks above the diagonal the new version performed
    better. Coverage 1.00 means all branchings are covered.}
  \label{pic:tc24_vs_tc25}
\end{wrapfigure}
\addtolength{\abovecaptionskip}{12mm}

We show the effectiveness of the presented algorithm empirically. We consider
two versions of the gray-box fuzzer \fizzer. The original
version~\cite{fizzer.24.testcomp} participated in
Test-Comp~2024~\cite{testcomp_results}. The binary is available on
Zenodo~\cite{fizzer_zenodo_testcomp}. The new version, comprising the presented
algorithm, participated in Test-Comp~2025~\cite{testcomp_results_2025}. The
binary of this version is also available on
Zenodo~\cite{fizzer_zenodo_testcomp_2025}.

We evaluated both versions of \fizzer on all benchmarks from Test-Comp
2024~\cite{testcomp_benchmarks_2024} used in the category Cover-Branches. This
benchmark suite consists of 9825 programs. The time reserved for analysis of
each benchmark was set to 5 minutes. And 5GB was the memory limit.

For each benchmark program each version of \fizzer generated a set of inputs to
the program. By executing the program for all the inputs we obtain the count of
\abe{}s covered (see Remark~\ref{rem:copverage}) in the program. Clearly, the
version of \fizzer covering more \abe{}s performs better on the program. In our
evaluation, we measured the coverage of \abe{}s indirectly. We used a tool
TestCov~\cite{testcov_repo} which is an official tool of Test-Comp for measuring
branch coverage of sets of inputs generated by competing tools. The branch
coverage is the coverage of branchings between basic blocks. We know that
evaluation of most \abe{}s have impact on the branchings. Therefore, there is
strong correlation between coverage of \abe{}s and program branchings.

The results from the comparison of both version are presented in
Fig.~\ref{pic:tc24_vs_tc25}. We see more dots above the diagonal, meaning the
new version of \fizzer performs better. On average, the new version achieved the
coverage 66.5\% per benchmark while the original version 59.8\%.


\section{Related Work}
\label{sec:related}

Gray-box fuzzers typically collect information (via probes instrumented to the
analyzed program) about a path taken during program's execution. The
representation of the path differs across fuzzing approaches. For instance,
\textsc{AFL}~\cite{AFL13} collects counts of executed transitions between basic
blocks. \textsc{libFuzzer}~\cite{libFuzzer} uses
\textsc{SanitizerCoverage}~\cite{SanitizerCoverage} for instrumentation. The
probes provide monitoring of transitions between basic blocks, basic blocks
entries, or function entries. In addition, the sanitizer can also generate a
table of control-flow in a function. \textsc{Angora}~\cite{angora18} collects
information about executed transitions between basic blocks together with a hash
of the callstack. Each call site is associated with a randomly chosen ID. The
call stack is modelled by a list of call site IDs. The hash is then computed as
XOR of all IDs in the stack. \textsc{VUzzer}~\cite{VUzzer17} monitors basic
blocks entries and also comparison instructions. \textsc{MaxAFL}~\cite{MaxAFL21}
monitors transitions between basic blocks similarly to \textsc{AFL}. In
addition, there are monitored comparison instructions. The original
\fizzer~\cite{fizzer.24.tacas} monitors \abe{}s together with call-stack hashes
computed similarly to \textsc{Angora}.

We defined the coverage problem for monitoring of \abe{}s. This is directly
relevant for fuzzers monitoring comparison instructions. Fuzzers monitoring
transitions between basic blocks or basic blocks entries can use an adopted
definition, where the black-bock functions comprise evaluation of all \abe{}s
involved in computation of whole branching conditions between basic blocks.

A taint-flow analysis is also popular in gray-box fuzzing, because it gives a
fuzzer a relation between the input and components (like \abe{}s or basic block
transitions) of corresponding execution path under reasonable computation costs.
Since dynamic taint-flow analysis typically leads to a heavy-weight
instrumentation, \textsc{Angora}~\cite{angora18} builds two versions of the
instrumented program; with and without taint tracking.
\textsc{MaxAFL}~\cite{MaxAFL21} in contrast applies a light-wight static taint
analysis and \textsc{VUzzer}~\cite{VUzzer17} applies both static and dynamic
taint-flow analysis. The original version of \fizzer~\cite{fizzer.24.tacas}
emulated a taint-flow analysis by analyzing execution results from one-bit input
mutations. The new version of \fizzer uses dynamic taint-flow analysis. However,
instead of instrumenting the program, individual instructions are interpreted;
the taint tracking is incorporated into the instruction's interpreter.

Regarding the coverage problem, the taint-flow analysis allows for inferring
parameters of the black-bock function comprised in the vector $\vec{F}$.
Therefore, the coverage problem is applicable to many fuzzing approaches either
directly or with minor modifications.

Bit-mutations and/or random mutations are common in
fuzzers~\cite{AFL13,MaxAFL21,libFuzzer,VUzzer17,fizzer.24.tacas}. Here we
present their more effective versions; they are performed in a local space. The
key advantage is that constraints along the path up to the considered one are
automatically taken into the account. Lastly, the gradient descent is also used
in fuzzing~\cite{angora18,MaxAFL21,SLF19,matryoshka19,neuzz19,fizzer.24.tacas}.
Here we showed how this approach can be optimized by performing it in a local
space.

\section{Conclusion}
\label{sec:conclusion}

We formally defined the coverage problem for gray-box fuzzing, where probes
instrumented to the analyzed program record information about evaluation of
\abe{}s and signed distances to the opposite evaluations in the executed program
path. We also described an effective algorithm for finding a solution of a given
coverage problem. We showed the effectiveness of the algorithm empirically on
benchmark suite of TestComp 2024, where we compared achieved results with those
from the original version of the tool, into which we implemented the algorithm.

\bibliographystyle{splncs04}
\bibliography{fizzer_local_search}

\end{document}